\documentclass[preprint,amsmath,amssymb,aps,superscriptaddress]{revtex4-1}

\usepackage{graphicx}
\usepackage{dcolumn}
\usepackage{bm}
\usepackage{color}
\usepackage{multirow}
\usepackage{epstopdf}

\begin{document}

\preprint{APS/123-QED}

\title{Impedance Matching of Atomic Thermal Interfaces Using Primitive Block Decomposition}

\author{Carlos A. Polanco}
\email{cap3fe@virginia.edu}
\affiliation{Department of Electrical and Computer Engineering, University of Virginia, Charlottesville,VA-22904.}%

\author{Christopher B. Saltonstall}
\author{Pamela M. Norris}
\author{Patrick E. Hopkins}
\affiliation{Department of Mechanical and Aerospace Engineering, University of Virginia, Charlottesville,VA-22904.}%

\author{Avik W. Ghosh}
\email{ag7rq@virginia.edu}
\affiliation{Department of Electrical and Computer Engineering, University of Virginia, Charlottesville,VA-22904.}%

\date{\today}

\begin{abstract}
We explore the physics of thermal impedance matching at the interface between two dissimilar materials by controlling the properties of a single atomic mass or bond. The maximum thermal current is transmitted between the materials when we are able to decompose the entire heterostructure solely in terms of primitive building blocks of the individual materials. Using this approach, we show that the minimum interfacial thermal resistance arises when the interfacial atomic mass is the arithmetic mean, while the interfacial spring constant is the harmonic mean of its neighbors. The contact induced broadening matrix for the local vibronic spectrum, obtained from the self-energy matrices, generalizes the concept of acoustic impedance to the nonlinear phonon dispersion or the short-wavelength (atomic) limit. 
\end{abstract}

\pacs{Valid PACS appear here}
\maketitle


\section{Introduction}

Today's experimental techniques are opening up the possibility of tuning thermal conductivity of materials by engineering their thermal impedance at the nanoscale \cite{Cahill}. At these characteristic lengths ($\sim$10nm), thermal boundary conductance (TBC) of interfaces provide a major contribution to the thermal conductance of the system, making critical the understanding of impedance matching at interfaces. Phonon transport across an interface is a convoluted process that involves the differing phonon modes, the short coherence lengths of the quantized vibrations and their broadband transport properties. Also, it involves complex and diverse interfacial atomic structures that depend strongly on materials and fabrication protocols. Several experiments \cite{Hopkins08,Hopkins10,Hopkins11_1,Hopkins11_2,Hopkins11_3,Hsieh11,Chen,Duda12} and simulations \cite{Hu09,Hu10,Ong10,Duda11,Shen11,wang11} have already shown the dependence of TBC with interfacial impurities, mixing, defects, chemistry or bond strength. Nevertheless, the standard models to calculate TBC, the acoustic mismatch model \cite{AMM} and the diffuse mismatch model \cite{DMM}, completely neglect the properties of the interface. Although some work has been done to include those properties into a model \cite{Lumpkin,Prasher,Bli,Hopkins12,Duda12}, a proper identification of the key physics determining TBC is still incomplete but it is crucial for impedance matching design at the nanoscale. This will lead the emerging field of phonon engineering to follow the successful steps of electronics and photonics, where engineering of nanoscale properties has endowed the fields with high degrees of tunability.

\begin{figure}[tb]
	\centering
	\includegraphics[width=86mm]{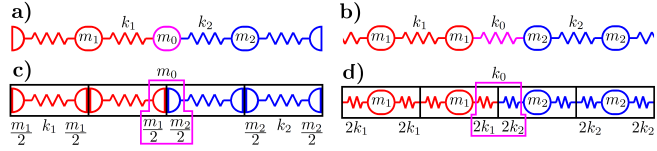}
	\caption{{\bf a)} 1D interface between dissimilar materials with an arbitrary atom in between. {\bf b)} Optimal coupling between the contacts happens when we can describe the entire heterostructure only in terms of building blocks of individual materials. This decomposition makes the optimal atomic mass the arithmetic mean of its neighbors $m_0=m_1/2+m_2/2$. {\bf c)} Interface with an arbitrary bond in between. {\bf d)} Maximum thermal conductance occurs when the interfacial spring constant is the harmonic mean of its neighbors $1/k_0=1/2k_1+1/2k_2$ (recall that a spring half as short is twice as strong), which follows again from a decomposition in terms of blocks of individual materials.}
	\label{figsi}
\end{figure}

While the overall goal of our study is to explore the physics of thermal impedance matching at interfaces covering the entire gamut from 1D to 3D, from linear to non linear dispersion and from coherent to incoherent transport, we will start building our intuition by studying coherent thermal impedance matching between two dissimilar 1D materials by controlling the properties of a single mass (Fig.~\ref{figsi}{\bf a}) or spring (Fig.~\ref{figsi}{\bf b}) in between. This toy model presents a starting point to understand ballistic contributions to TBC by important factors already identified in the literature, like interfacial impurities, mixing, defects, chemistry or bond strength \cite{Hopkins08,Prasher,Hopkins10,Hopkins11_1,Hopkins11_2,Hopkins11_3,Chen,Duda12}. In fact, some authors have used this toy model to support their Molecular Dynamic simulation results arguing the increase of TBC with increase of bond strength \cite{Hu09,Hu10,Shen11}.

The results for coherent, 1-D thermal impedance matching are incredibly diverse. For example, to achieve maximum thermal conductance we want the interfacial impedance to maximize the area under the transmission function, like a {\it broadband} filter. Following this criterion, we point out that the best matching interfacial mass ($m_0$) for the single mass junction (Fig.~\ref{figsi}{\bf a}) is the {\it arithmetic} mean between the masses of the contacts \cite{chris}. For a single spring junction (Fig.~\ref{figsi}{\bf b}), Zhang et al. \cite{Bli} found that the best matching interfacial spring constant ($k_0$) is the {\it harmonic} mean between the contact springs. When the goal is to achieve maximum phonon transmission around a fixed frequency, our expectations based on our knowledge of optical antireflection coatings posit that unity phonon transmission would require a quarter wave plate with an impedance equal to the geometric mean of its neighbors. In view of these diverse results, our aim here is to put these averages on a common footing and motivate them qualitatively in terms of the intrinsic physical properties of the junction itself.

The central point of this paper is that the degree of mismatch at a single atom or bond interface depends on our ability to express the entire heterostructure solely in terms of primitive building blocks on either side. For instance, we find that the optimal mass (Fig.~\ref{figsi}{\bf a}) is one that can be decomposed precisely into two half-masses arising from the materials on either side (Fig.~\ref{figsi}{\bf c}). This decomposition makes the optimal mass the {\it arithmetic} mean of its neighbors, i.e. $m_0=m_1/2+m_2/2$. For an analogous decomposition of the spring constant (Fig.~\ref{figsi}{\bf b} and {\bf d}), we find that the optimal spring constant equals the {\it harmonic} mean of its neighbors, i.e. $1/k_0=1/2k_1+1/2k_2$ (Recall that springs in series add like resistances in parallel). Any deviation from those optimal decompositions (``Optimally Coupled Interfaces"-OCI) adds an extra barrier for heat carriers reducing the interfacial transmission. 

The thermal conductance for OCI is characterized by the contact induced broadening matrix $\Gamma(\omega)$ extracted from the local vibronic spectrum, which generalizes the concept of acoustic impedance ($Z$). $\Gamma$ not only includes non linear dispersion and short-wavelength (atomic) limit effects but its matrix character can account for the different modes or channels available for transport when higher dimensions are considered. Also, this character can include intricate chemical details at the interface, which may greatly affect the transport process as shown recently by Losego et~at. \cite{cahill2012}. It is worth emphasizing that $\Gamma$ alone is not enough to correctly represent general phonon transport. The broadening's Hilbert transform must be also included in the Green's function to properly account for the sum rule of the local density of states \cite{Datta}.

This generalization not only relates the continuum formalism \cite{Pain} with the discrete Non-Equilibrium Green's Functions (NEGF) formalism \cite{Datta,Hopkins09,Mingo03,zhang1D,Wang}, but also provides a way to extrapolate known results based on acoustic impedance to OCI. For instance, we can totally eliminate the interfacial reflection by impedance matching of the $\Gamma$ matrices (more precisely, the projected self-energies, $\Sigma$), realized when $\Gamma$ for the central layer equals the geometric mean of its neighbors. The generalization may also allow us to use existing techniques from other engineering fields in phonon engineering. For instance, broadband filter techniques from microwave engineering may be useful to engineer interfaces with maximum thermal conductance.

The document begins by explaining the idea of splitting 1D chains into primitive blocks, which define the properties of contacts or semi-infinite chains (Section \ref{secblockcontacts}). Then, using the block concept, phonon transmission is calculated in section \ref{sectransmission}, where it is also shown that maximum thermal conductivity occurs when the entire heterostructure can be expressed solely in terms of the building blocks on either side of the interface. These types of interfaces (OCI) are studied and characterized in section \ref{secpci}, where it is shown that OCI generalizes an abrupt interface in the continuum limit with $\Gamma$ generalizing $Z$.

\section{Block Partition of 1D Chain and Contacts}\label{secblockcontacts}

\begin{figure}[t]
	\centering
	\includegraphics[width=74mm]{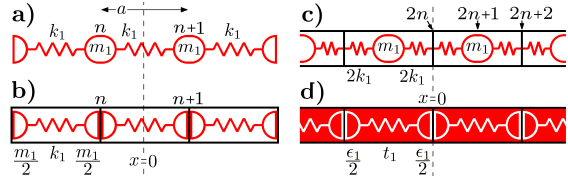}	
	\caption{{\bf a)} 1D infinite chain of masses $m_1$, separated by $a$, joined by springs with force constant $k_1$. {\bf b)} and {\bf c)} Same chain separated into different blocks whose boundaries define different contacts. {\bf b)} Partition into half-mass blocks with half masses across the boundary. {\bf c)} Partition into half-spring  blocks with half springs across the boundary. {\bf d)} The half-spring blocks can be reinterpreted as half-mass blocks, provided the corresponding mass and spring constant are frequency dependent. The non-white background of the blocks represent this dependence.}
	\label{fig1Dchains}
\end{figure}

An infinite 1D chain of masses coupled by springs (Fig.~\ref{fig1Dchains}a) can be decomposed into different arrays of primitive blocks (Figs. \ref{fig1Dchains}b and \ref{fig1Dchains}c ). According to the blocks, different contacts, i.e. semi-infinite chains, can be built from the same homogeneous material. As we will also show, we can equally view  a chain of half-spring blocks as a virtual chain of half-mass blocks, provided that the corresponding mass and spring constant are {\it frequency dependent}.

Consider a 1D infinite chain of masses $m_1$, separated by a distance $a$ and connected by springs with force constant $k_1$ (Fig.~\ref{fig1Dchains}{\bf a}). Newton's equation for the normalized displacement of the $n^{th}$ atom, $\mu_n(t)=u_ne^{-i\omega t}$ with dimensions inverse square root of mass, i.e. $[\mu]=\left[M^{-1/2}\right]$,  is given by
\begin{equation}
\omega^2m_1u_{n}=-k_1u_{n-1}+2k_1u_n-k_1u_{n+1}.
\label{equ1dinfchain}
\end{equation}
This set of periodic equations is solved by plane waves $u_n=Ae^{iq_1na}$, satisfying the dispersion relation
\begin{equation}
\omega^2m_1=2k_1-2k_1\cos(q_1a).
\label{equdr1Dchain}
\end{equation}
Splitting each mass into its series equivalent $m_1=m_{1}/2+m_{1}/2$, the chain can be partitioned into blocks with boundaries at the {\it masses}, i.e. half-mass blocks (Fig.~\ref{fig1Dchains}{\bf b}). In this case, Eq.~\ref{equ1dinfchain} and \ref{equdr1Dchain} can be reorganized to reflect the partition as
\begin{equation}
\omega^2\left(\frac{m_1}{2}+\frac{m_1}{2}\right)u_n=-k_1u_{n-1}+2k_1u_n-k_1u_{n+1}
\label{equ1dinfsmb}
\end{equation}
\begin{equation}
\omega^2\left(\frac{m_1}{2}+\frac{m_1}{2}\right)=2k_1-2k_1\cos(q_1a).
\label{equdrsm}
\end{equation}
Note that the plane waves solving those equations represent $|A|^2Nm_1$ propagating phonons of energy $\hbar\omega$ ($N$ is the number of atoms in the chain) and carry a thermal current given by 
\begin{equation}
J=\hbar k_1\sin(q_1a)|A|^2=\hbar\frac{\Gamma_1^{hm}}{2}|A|^2=\hbar\omega\frac{m_1}{a}v_g(\omega)|A|^2,
\label{equtcbypwsm}
\end{equation}
where
\begin{equation}
\Gamma_1^{hm}=2k_1\sin(q_1a)
\label{equGammasm}
\end{equation}
(``hm'' stands for half-mass) is the non-zero entry of the broadening matrix used in NEGF formalism and $v_g(\omega)$ is the frequency dependent phonon group velocity.

Similarly, we can split each spring into its series equivalent $1/k_1=1/2k_{1}+1/2k_{1}$, separating the chain (Fig.~\ref{fig1Dchains}{\bf a}) into blocks with boundaries at the {\it springs}, i.e. half-spring blocks (Fig.~\ref{fig1Dchains}{\bf c}). The latter system is described by
\begin{equation}
\omega^2m_1u_{2n+1}=-2k_{1}u_{2n}+2(k_{1}+k_{1})u_{2n+1}-2k_{1}u_{2n+2}, 
\label{equssodd}
\end{equation}
\begin{equation}
0=-2k_{1}u_{2n-1}+2(k_{1}+k_{1})u_{2n}-2k_{1}u_{2n+1}. 
\label{equsseven}
\end{equation}
Solving for $u_{2n}$ and $u_{2n+2}$ from Eq.~\ref{equsseven} and replacing into Eq.~\ref{equssodd} yields Eq.~\ref{equ1dinfchain}. More interestingly, solving for $u_{2n-1}$ and $u_{2n+1}$ from Eq.~\ref{equssodd} and replacing into Eq.~\ref{equsseven}, results in an equation similar to Eq.~\ref{equ1dinfsmb}
\begin{equation}
\omega^2\left(\frac{\epsilon_1}{2}+\frac{\epsilon_1}{2}\right)u_{2n}=-t_1u_{2n-2}+2t_1u_{2n}-t_1u_{2n+2},
\label{equ1dinfssb}
\end{equation}
where 
$$\epsilon_1=\frac{m_1}{1-\frac{\omega^2}{\omega_{c1}^2}},\ \ \ \ \ t_1=\frac{k_1}{1-\frac{\omega^2}{\omega_{c1}^2}}$$
are {\it frequency dependent coefficients} and the cut off frequency is given by $\omega_{c1}=2\sqrt{k_1/m_1}$. In other words, the half-spring block chain can be interpreted as a virtual half-mass block chain having frequency dependent masses and springs. 

This analogy permits the extrapolation of algebraic treatments, like NEGF, from half-mass block to virtual half-mass block chains. For instance, plane waves describing the displacement at the boundaries of half-spring blocks $u_{2n}=Ae^{iq_12n\frac{a}{2}}=Ae^{iq_1na}$ satisfy the dispersion relation 
\begin{equation}
\omega^2\left(\frac{\epsilon_1}{2}+\frac{\epsilon_1}{2}\right)=2t_1-2t_1\cos(q_1a)
\label{equdrss}
\end{equation}
and carry a thermal current 
\begin{equation}
J=\hbar t_1sin(q_1a)|A|^2=\hbar\frac{\Gamma_1^{hs}}{2}|A|^2=\hbar\omega\frac{\epsilon_1}{a}v_g(\omega)|A|^2
\label{equtcbypwss}
\end{equation}
with 
\begin{equation}
\Gamma_1^{hs}=2t_1\sin(q_1a)
\label{equGammass}
\end{equation}
(``hs'' stands for half-spring) the non-zero entry of the broadening matrix used in NEGF formalism for the virtual chain.

Although the same infinite chain or bulk material can be built from any block, different contacts are created from different blocks. Indeed, the block choice defines the edge of the contact, the positions in space described by displacement plane waves $Ae^{iqna}$ (block boundaries) and more importantly the thermal current carried by those waves. One striking example of the difference between half-mass and half-spring contacts arises when we connect them together. A phonon impinging on such an interface has non-zero probability of reflection, unlike a phonon propagating in a single block chain. Note that this interface mimics a growth defect in a 1D crystal.

\section{Transmission Using Blocks}\label{sectransmission}

Since a set of phonons of equal energy propagating in a crystal are well represented by plane waves, the transmission probability of phonons impinging at an interface can be calculated from the ratio between the thermal currents carried by the transmitted and incident waves. This section presents phonon transmission calculations using the block concept to simplify the process. It is shown that maximum transmission at every frequency, and therefore maximum thermal conductance, happens  when the entire heterostructure can be expressed solely in terms of building blocks on either side. This idea is equivalent to choosing the interfacial atomic mass as the arithmetic mean or the interfacial spring constant as the harmonic mean of its neighbors.

\subsection{Interface with Mass Junction}

\begin{figure}[t]
	\centering
	\includegraphics[width=86mm]{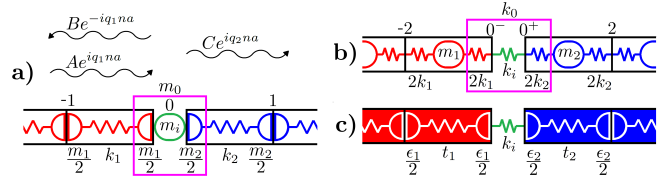}
	\caption{Decomposition of interfaces into blocks for transmission calculations. An upper bound for transmission happens when the impurity atom $m_i$ or bond $1/k_i$ are zero. {\bf a)} Atom junction interface split into half-mass blocks. ({\bf b}) Bond junction interface split into half-spring  or ({\bf c}) virtual half-mass blocks. In this case $u_{2n} = Ae^{iq_1na} + Be^{iq_1na}$.}
	\label{wave_F}
\end{figure}

Imagine chopping the materials of Fig.~\ref{figsi}{\bf a} into half-mass blocks and the interfacial mass into a series equivalent that completes the contacts' blocks plus some residual mass $m_i$, i.e.  
\begin{equation}
m_0=\frac{m_1}{2}+m_i+\frac{m_2}{2}
\label{equmi}
\end{equation}
(Fig.~\ref{wave_F}{\bf a}). Assuming incident, reflected and transmitted plane wave solutions,
transmission is given by the ratio of transmitted over incident thermal currents 
$$T=\frac{J_t}{J_i}=\frac{\Gamma_2^{hm}}{\Gamma_1^{hm}}\left|\frac{C}{A}\right|^2.$$
The relationship between $A$ and $C$ is found by substituting the assumed solution
\begin{equation}
u_n=\begin{cases}
Ae^{iq_1na}+Be^{-iq_1na} & n\leq0 \\
Ce^{iq_2na} & n\geq0
\end{cases}
\end{equation}
into Newton's equation at the interface of Fig.~\ref{wave_F}{\bf a} $(n=0)$
\begin{equation}
\omega^2\left(\frac{m_1}{2}+m_i+\frac{m_2}{2}\right)u_{0}=-k_1u_{-1}+(k_1+k_2)u_0-k_2u_{1}.
\label{equsm0}
\end{equation}
This process is simplified using Eq.~\ref{equdrsm}, noting that the real part of the right hand side of Eq.~\ref{equsm0} exactly cancels $\omega^2(m_1+m_2)u_0/2$, which yields
\begin{equation}
\omega^2m_iu_0=i(A-B)\frac{\Gamma_1^{hm}}{2}-iC\frac{\Gamma_2^{hm}}{2}.
\label{equ2interfacesm}
\end{equation}
Combining this result with the fact that $u_0=A+B=C$, the transmission $T$ is found to be a Breit-Wigner form
\begin{equation}
T(\omega,m_i)=\frac{4\Gamma_1^{hm}\Gamma_2^{hm}}{4\omega^4m_i^2+\left(\Gamma_1^{hm}+\Gamma_2^{hm}\right)^2},
\label{equTsm}
\end{equation}
with $m_i$ being the deviation of the interfacial mass $m_0$ from the arithmetic mean between the contact masses (Eq.~\ref{equmi}). 

Note that the largest possible transmission for every $\omega$ is obtained when $m_i=0$. This choice maximizes the thermal current flowing across the interface and the thermal conductance of the system. That is,
\begin{align*}
I_{0}&=\int d\omega\frac{\hbar\omega}{2\pi}T(\omega,0)(N_1-N_2) \\
&\geq \int d\omega\frac{\hbar\omega}{2\pi}T(\omega,m_i)(N_1-N_2)=I_{m_i}.
\end{align*}
In this particular case, the system becomes equivalent to an abrupt interface between contacts built with half-mass blocks, which is referred as an ``Optimally Coupled Interface" (OCI). When $m_i\neq0$, transmission decreases (for all $\omega$) so $m_i$ can be associated with an extra barrier lowering the thermal conductance. Specifically a delta scattering center like a single point impurity or defect at the interface.

\subsection{Interface with Spring Junction}

Imagine now chopping the contacts of Fig.~\ref{figsi}{\bf b} into half-spring blocks and the interfacial spring into its series equivalent
\begin{equation}
\frac{1}{k_0}=\frac{1}{2k_1}+\frac{1}{k_i}+\frac{1}{2k_2}
\label{equki}
\end{equation}
 (Fig.~\ref{wave_F}{\bf b}). Assuming incident, reflected and transmitted plane wave solutions at the blocks' boundaries, transmission is given by the ratio of transmitted over incident thermal currents 
$$T=\frac{J_t}{J_i}=\frac{\Gamma_2^{hs}}{\Gamma_1^{hs}}\left|\frac{C}{A}\right|^2.$$
The relationship between $A$ and $C$ is found by substituting the assumed solution
\begin{equation}
u_{2n}=\begin{cases}
Ae^{iq_1na}+Be^{-iq_1na} & n<0 \text{ and } n=0^- \\
Ce^{iq_2na} & n>0 \text{ and } n=0^+
\end{cases}
\end{equation}
into Newton's equation at the interface ($n=0^-$ and $n=0^+$) for the virtual chain (Fig.~\ref{wave_F}{\bf c})
\begin{equation}
\omega^2\frac{\epsilon_1}{2}u_{0^-}=-t_1u_{-2}+(t_1+k_i)u_{0^-}-k_iu_{0^+}
\label{equss0m}
\end{equation}
\begin{equation}
\omega^2\frac{\epsilon_2}{2}u_{0^+}=-k_iu_{0^-}+(k_i+t_2)u_{0^+}-t_2u_{2}.
\label{equss0p}
\end{equation}
This process is simplified using Eq.~\ref{equdrss}, noting that the real part of $t_1(u_{0^-}-u_{-2})$ from Eq.~\ref{equss0m} and $t_2(u_{0^+}-u_{2})$ from Eq.~\ref{equss0p} exactly cancel $\omega^2\epsilon_1u_{0^-}/2$ and $\omega^2\epsilon_2u_{0^+}/2$ respectively, which yields
\begin{equation}
0=k_iu_{0^-}-k_iu_{0+}+i(A-B)\frac{\Gamma_1^{hs}}{2}
\label{equss_1}
\end{equation}
and
\begin{equation}
0=-k_iu_{0^-}+k_iu_{0+}-iC\frac{\Gamma_2^{hs}}{2}.
\label{equss_2}
\end{equation}
Combining these two results with $u_{0^-}=A+B$ and $u_{0^+}=C$, the transmission $T$ is given by
\begin{equation}
T(\omega,k_i^{-1})=\frac{4\Gamma_1^{hs}\Gamma_2^{hs}}{\frac{1}{4k_i^2}\left(\Gamma_1^{hs}\Gamma_2^{hs}\right)^2+\left(\Gamma_1^{hs}+\Gamma_2^{hs}\right)^2},
\label{equTss}
\end{equation}
with $k_i^{-1}$ measuring the deviation of the interfacial spring $k_0$ from the harmonic mean between the contact springs (Eq.~\ref{equki}). 

Note that the largest possible transmission for every $\omega$ is obtained when $k_i^{-1}=0$. This choice maximizes the thermal current flowing across the interface and the thermal conductance of the system. In that case, the system also becomes equivalent to an abrupt interface between contacts built with half-spring blocks, which is also referred as an ``Optimally Coupled Interface". When $k_i^{-1}\neq0$, transmission (for all $\omega$) and thermal conductance decrease, so that $k_i^{-1}$ can be associated with an extra barrier at the interface.

\section{Optimally Coupled Interfaces}\label{secpci}

An {\bf ``Optimally Coupled Interface''} (OCI) is an abrupt interface between half-mass or half-spring block contacts, which was proven equivalent to the single atomic or bonding interface with maximum possible transmission or thermal conductance (Section \ref{sectransmission}). This section shows that an OCI can be thought of as a step barrier for phonons responsible for the scattering due to a change in propagation medium. On the other hand, a non-OCI is represented by the same step barrier plus an extra barrier caused by a deviation from the optimal case ($m_i\neq0$ or $k_i^{-1}\neq0$). The extra barrier decreases thermal conductance and can be associated with additional scattering mechanisms at the interface, such as impurities, mixing or dislocations. This section presents a useful way to visualize transmission in OCI from the contact broadenings and extends the concept of OCI to abrupt junctions between contacts built with different types of blocks.

The section also shows that OCI generalizes an abrupt interface in the continuum limit without the long wavelength constraint. Moreover, the {\it bulk} property $Z$ (acoustic impedance) is generalized by the {\it contact} property $\Gamma$ (broadening), which unlike $Z$ includes the atomistic details of the contact's edge and the non-linear effects of phonon dispersion. This analogy shows a way to extrapolate previous results for interfaces in the continuum limit to the discrete limit by replacing abrupt interfaces with OCI and $Z$ with $\Gamma$. For instance, the result of a thermal antireflection coating for a quarter wave length plate is obtained when the plate broadening equals the geometric mean of the individual contact broadenings.

\subsection{Continuous vs. Discrete Limit}\label{subseccvsd}

The continuous medium approximation assumes that the wavelengths of interest are large enough ($\lambda>>a$) so that the atomistic details of the media are ignorable, the dispersion is linear and the group velocity is constant. Within this approximation, the scattering problem at an interface (Table \ref{tabparallel}-{\bf a}) is solved by assuming incident, reflected and transmitted plane waves solutions and imposing boundary conditions on them to guarantee the validity of the wave equation at the interface. These conditions are nicely simplified introducing the concept of acoustic impedance (Table \ref{tabparallel}-{\bf b} and {\bf c} \cite{Pain})
\begin{equation}
Z=\rho v_g=\frac{m}{a}\left(a\sqrt{\frac{k}{m}}\right)=\sqrt{km}.
\label{equimp}
\end{equation}
From them, the ratio of the wave amplitudes is calculated and then transmission is found from the ratio of transmitted over incident energy currents (Table \ref{tabparallel}-{\bf e}). Scattering at these interfaces can be connected with {\it medium} mismatch using the reflection coefficient 
\begin{equation}
R=1-T=\left(\frac{Z_1-Z_2}{Z_1+Z_2}\right)^2,
\end{equation}
which vanishes only when $Z_1|A|^2=Z_2|A|^2$. In words, if plane waves of the same amplitude do not carry the same energy current in both {\it media} then some energy has to be reflected. That is, the scattering is solely caused by mismatch of the medium properties.
\begin{table}[htb]
\centering
\begin{tabular}{c|c|c|}
\cline{2-3}
& {\bf Continuous} & {\bf Discrete} \\ 
\multirow{1}{*}{\bf{a}} & \includegraphics[width=35mm]{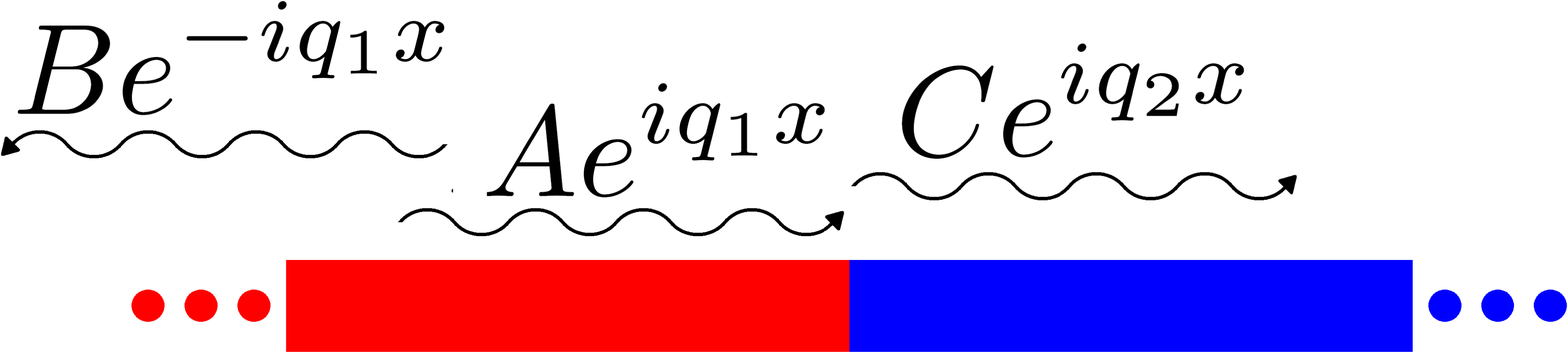} & \includegraphics[width=35mm]{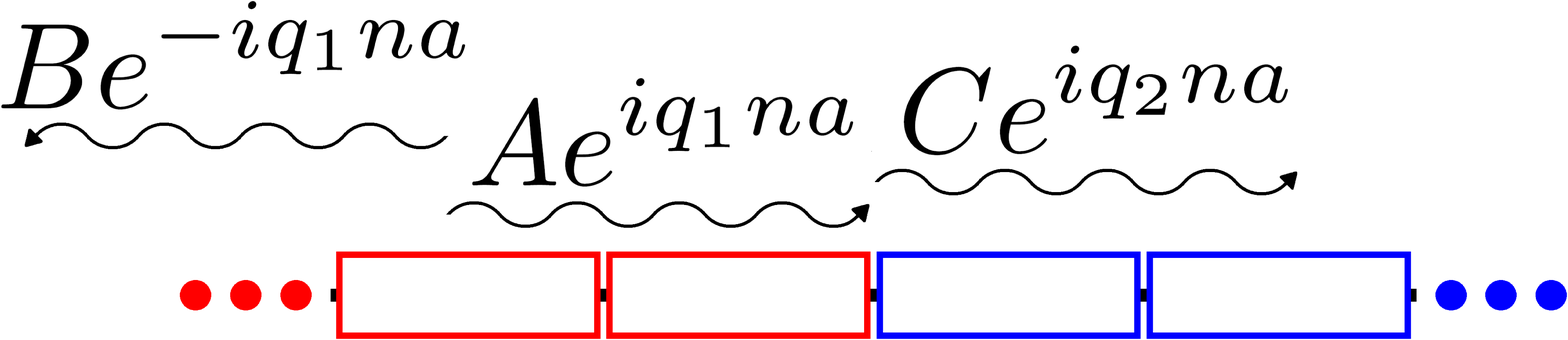} \\ \cline{2-3}
& Acoustic Impedance & Broadening \\ \cline{2-3}
\bf{b} & $Z=\rho v_g$ & $\Gamma(\omega)=2\omega\rho(\omega) v_g(\omega)$ \\ \cline{2-3}
& \multicolumn{2}{|c|}{From Boundary Conditions} \\ \cline{2-3}
\multirow{2}{*}{\bf{c}} & $A+B=C$ & $A+B=C$ \\
& $Z_1(A-B)=Z_2C$ & $\Gamma_1(A-B)=\Gamma_2C$ \\ \cline{2-3}
& \multicolumn{2}{|c|}{Energy Current} \\ \cline{2-3}
\bf{d} & $J_A\propto Z|A|^2$ & $J_A\propto \Gamma|A|^2$ \\ \cline{2-3}
& \multicolumn{2}{|c|}{Transmission} \\ \cline{2-3}
\bf{e} & $T=\frac{4Z_1Z_2}{(Z_1+Z_2)^2}$ &  $T=\frac{4\Gamma_1\Gamma_2}{\left(\Gamma_1+\Gamma_2\right)^2}$ \\ \cline{2-3}
\end{tabular}
\caption{Parallel between an abrupt interface in the continuum limit and an Optimally Coupled Interface. The resembling suggests that the contact induced broadening matrix $\Gamma$ generalizes the concept of acoustic impedance ($Z$) to the nonlinear phonon dispersion as well as the short-wavelength (atomic) limit. For $Z$, $\rho=m/a$. For $\Gamma$, $\rho=m/a$ or $\rho=\epsilon/a$ according to the block choice. Note that the matrix character of $\Gamma$ can account for all the conduction modes available in higher dimensions, for interactions beyond first neighbor and for tensorial properties of materials.}
\label{tabparallel}
\end{table}

When the long wavelength constraint is relaxed, the frequency dependent group velocity, the cut-off frequency and the atomistic details of the interface affect the transmission (All included in calculations in section \ref{sectransmission}). Nevertheless, particularizing the transmission calculation in section \ref{sectransmission} to the optimal case ($m_i=0$ for mass junction and $k_i^{-1}=0$  for spring junction) displays the resemblance between an OCI and an abrupt interface in the continuum limit (Table \ref{tabparallel}). This suggests that OCI generalizes the continuous interface with $\Gamma$ playing the role of acoustic impedance $Z$. Note that in the long wavelength limit $\Gamma\rightarrow 2\omega Z$ (for both $\Gamma^{hm}$ and $\Gamma^{hs}$) and we recover the transmission result in terms of $Z$
\begin{equation}
T=\frac{4\Gamma_1\Gamma_2}{\left(\Gamma_1+\Gamma_2\right)^2}\xrightarrow{\lambda>>a} \frac{4Z_1Z_2}{(Z_1+Z_2)^2}.
\end{equation}

Similar to the continuum limit, scattering in OCI is solely due to {\it contact} mismatch. That is, if plane waves of the same amplitude do not carry the same energy current in both {\it contacts} then some energy has to be reflected. The subtle difference from {\it medium} to {\it contact} reflects the fact that unlike $Z$, $\Gamma$ is a {\it contact} property which ultimately depends on the block choice and carry information about the contact's edge. 

\begin{figure}[htb]
	\centering
	\includegraphics[width=70mm]{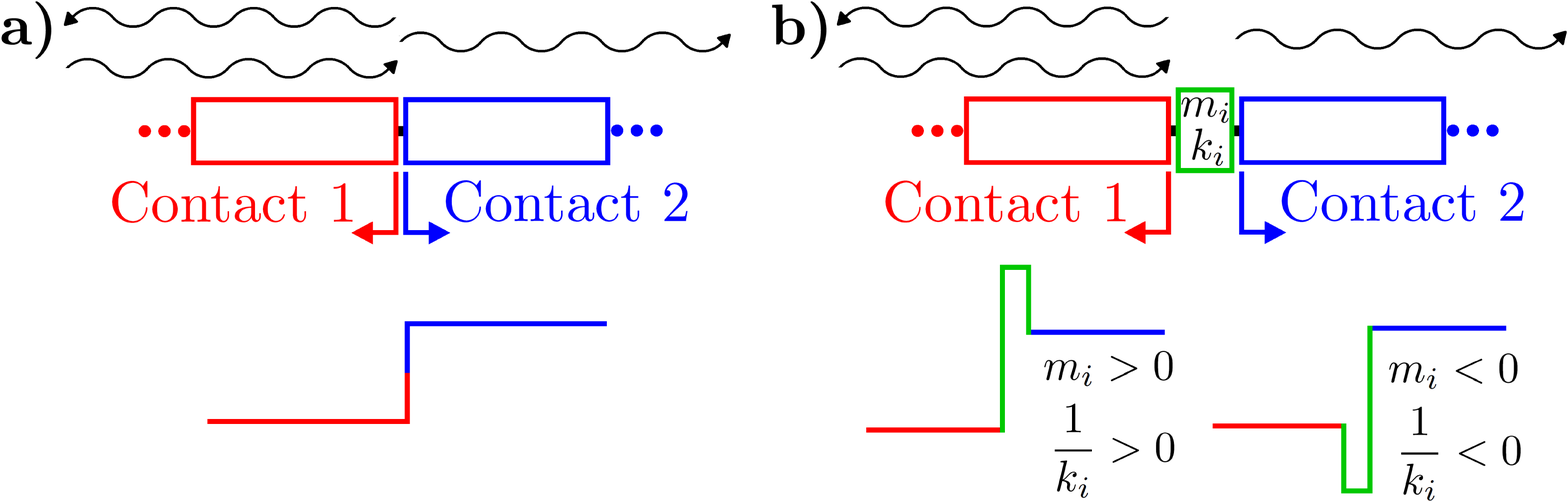}
	\caption{{\bf a)} Optimally Coupled Interface and its representation as a single barrier for phonons, which is responsible for the scattering due to propagating waves changing medium. {\bf b)} Non OCI and its representation as a step barrier plus an extra barrier caused by a deviation from the optimal case. This extra barrier decreases thermal conductance and can be associated with impurities, mixing or defects at the interface.}
	\label{figbarrier}
\end{figure}

After identifying contact mismatch scattering with the transmission functional defining OCI (Table \ref{tabparallel}-{\bf e}), the extra term decreasing the transmission in Eq.~\ref{equTsm} or \ref{equTss} is associated with an additional source of scattering at the interface. Following this train of ideas, an OCI is represented by a {\it frequency dependent} step barrier for phonons responsible for contact mismatch scattering while a non OCI is represented by the same step barrier plus an extra barrier that decreases transmission and can be associated with impurities, mixing or dislocations (Fig.~\ref{figbarrier}). 

Note that if non-symmetric blocks are used, the simplifications in Table \ref{tabparallel}-{\bf c} are not possible for every $\omega$. That is, abrupt interfaces between contacts built with non-symmetric  blocks do not resemble the equations in the continuum limit.

\subsection{Characteristics of OCI}\label{subseccharG}

OCI transmission can be visualized from the contact broadenings using the reflection coefficient,
\begin{equation}
R=1-T=\left|\frac{\Gamma_1-\Gamma_2}{\Gamma_1+\Gamma_2}\right|^2.
\label{equTpci}
\end{equation}
Unity transmission ($T(\omega_{*})=1$) is obtained when $\Gamma$'s match $(\Gamma_1(\omega_{*})=\Gamma_2(\omega_{*}))$ at a particular frequency $\omega_{*}$. That is, a phonon with energy $\hbar\omega_{*}$ does not see the interface. Null transmission ($T(\omega)=0$) is obtained if any of the $\Gamma$'s becomes 0 or imaginary. This defines a cut off frequency $\omega_c$ over which phonons do not propagate in the contact. Also, note that the frequency dependence of $\Gamma$ disallows the possibility of matching different contacts at every $\omega$, making scattering unavoidable. 

\begin{figure}[t]
	\centering
	\includegraphics[width=65mm]{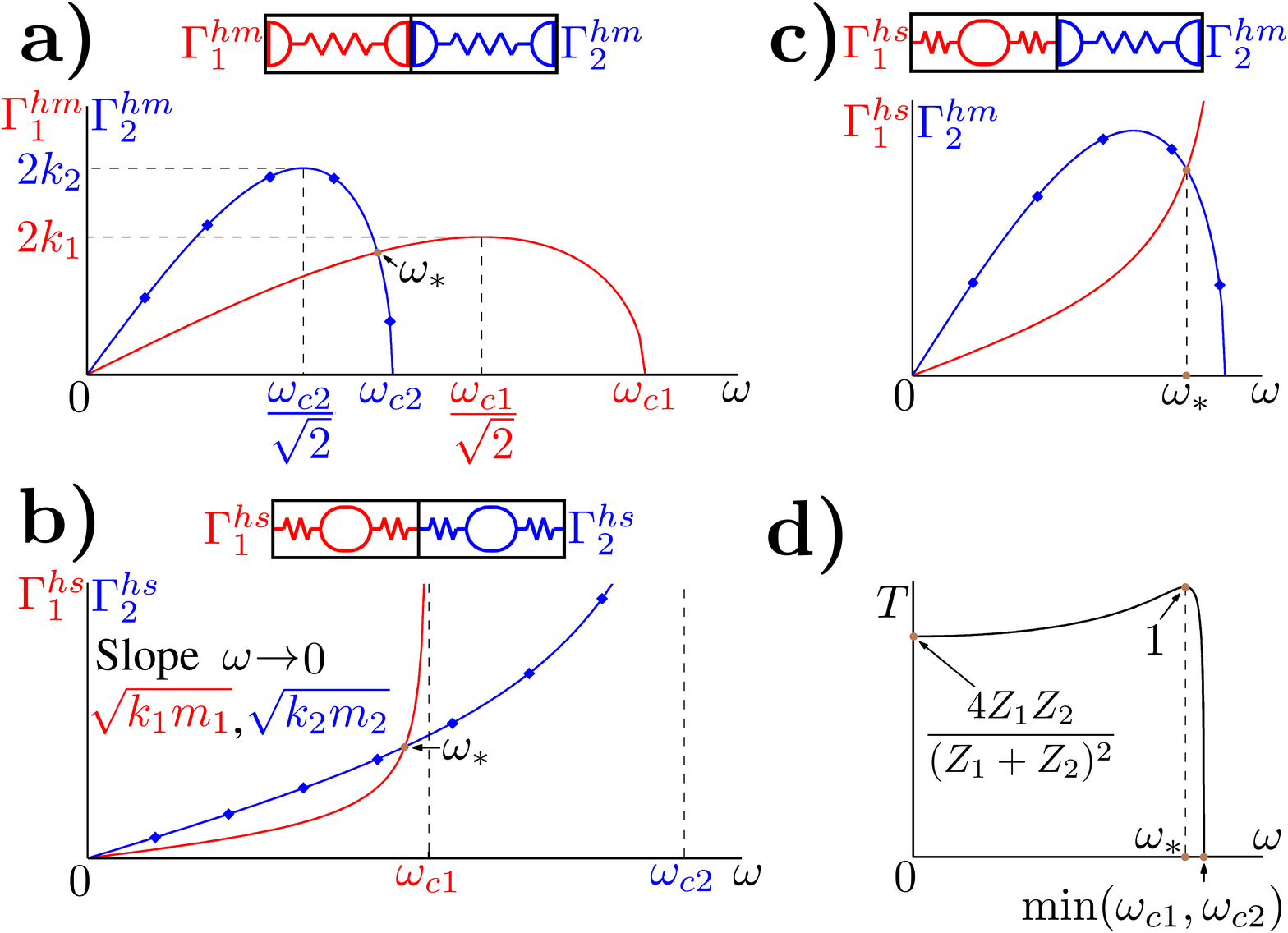}
	\caption{Transmission of OCI visualized from the broadening of the contacts ($\Gamma_1$ and $\Gamma_2$). {\bf a)}, {\bf b)} and {\bf c)} $\Gamma$'s for half-mass contacts, half-spring contacts and half-spring, half-mass contacts respectively. The dotted lines represent $\Gamma_2$. {\bf d)} Characteristic transmission function for all the cases before. At the particular frequency $\omega_{*}$, where $\Gamma$s intersect each other, transmission becomes unity and the materials match.} 
	\label{figGammasm}
\end{figure}

When a contact is built with {\it half-mass blocks}, $\Gamma_1$ from Eq.~\ref{equGammasm} can be rewritten using the dispersion relation as
\begin{equation}
\Gamma_1^{hm}=2k_1\sqrt{1-\left(1-\frac{2\omega^2}{\omega_{c1}^2}\right)^2}
\end{equation}
with $\omega_{c1}=2\sqrt{k_1/m_1}$. This concave function vanishes at $\omega=0$ and $\omega=\omega_{c_1}$ and has a maximum value $2k_1$ at frequency $\omega_{c_1}/\sqrt{2}$  (Fig.~\ref{figGammasm}a). With this function in mind, transmission of an OCI between half-mass block contacts (Fig.~\ref{figGammasm}d) can be visualized from a plot of the real part of both contact broadenings (Fig. \ref{figGammasm}a) and Eq.~\ref{equTpci}. Figs.~\ref{figGammasm}a and \ref{figGammasm}d show the case when the contacts {\bf match} at a particular frequency $\omega_{*}$, i.e. when $T(\omega_{*})=1$. This is only possible if $\Gamma$'s intersect, which requires $k_1<k_2$ and $\omega_{c1}>\omega_{c2}$ or $k_1>k_2$ and $\omega_{c1}<\omega_{c2}$. The intersection frequency $\omega_{*}$ is found by equating $\Gamma_1(\omega_{*})=\Gamma_2(\omega_{*})$ as
$$
 \omega_{*}^2=\begin{cases}
    \frac{4(m_1k_1-m_2k_2)}{m_1^2-m_2^2} & \text{if } m_1\neq m_2 \\
    [0,\min(\omega_{c_1},\omega_{c2})]         & \text{if } m_1=m_2 \text{ and } k_1=k_2	 \\
    \text{never}                                                 & \text{if } m_1=m_2 \text{ and } k_1\neq k_2
   \end{cases}
$$
Note that when acoustic impedances match the intersection frequency is zero, which makes sense since in the continuous regime any wave looks like a zero frequency wave. When the contacts {\bf do not match} or $\Gamma$'s never intersect, the transmission never reaches unity. This happens if $k_1>k_2$ and $\omega_{c1}>\omega_{c2}$ or $k_1<k_2$ and $\omega_{c1}<\omega_{c2}$. Another interesting case occurs when the cut-off frequencies are equal, which makes $\Gamma_1\propto\Gamma_2$ and therefore transmission is {\bf constant}.

When a contact is built with {\it half-springs blocks}, $\Gamma_1$ from Eq.~\ref{equGammass} can be rewritten using the dispertion relation as
\begin{equation}
\Gamma_1^{hs}=4k_1\sqrt{\frac{\omega^2}{\omega_{c_1}^2-\omega^2}}. 
\end{equation}
This convex function vanishes at $\omega=0$, is $\infty$ at $\omega=\omega_{c_1}$ and has a slope at $\omega=0$ of $2\sqrt{k_1 m_1}$ (Fig.~\ref{figGammasm}b). Similar to the former case, Fig.~\ref{figGammasm}b and d show the case when the contacts {\bf match}, which is only possible if $k_1m_1>k_2m_2$ and $\omega_{c1}>\omega_{c2}$ or $k_1m_1<k_2m_2$ and $\omega_{c1}<\omega_{c2}$. The intersection frequency $\omega_{*}$ is found equating $\Gamma_1(\omega_{*})=\Gamma_2(\omega_{*})$ as
$$
 \omega_{*}^2=\begin{cases}
    4\frac{k_1k_2}{m_1m_2}\frac{(m_1k_1-m_2k_2)}{k_1^2-k_2^2} & \text{if } k_1\neq k_2 \\
    [0,\min(\omega_{c_1},\omega_{c2})]         & \text{if } k_1=k_2 \text{ and } m_1=m_2	 \\
    \text{never}                                                 & \text{if } k_1=k_2 \text{ and } m_1\neq m_2
   \end{cases}
$$
The contacts {\bf do not match} if $k_1m_1<k_2m_2$ and $\omega_{c1}>\omega_{c2}$ or $k_1m_1>k_2m_2$ and $\omega_{c1}<\omega_{c2}$ and transmission is {\bf constant} if $\omega_{c1}=\omega_{c2}$.

\subsection{Other OCI}

\begin{figure}[htb]
	\centering
	\includegraphics[width=65mm]{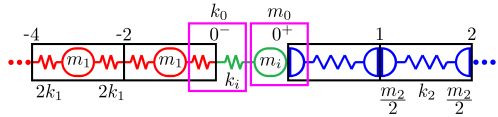}
	\caption{1D interface between dissimilar materials with an arbitrary bond and atom in between. An OCI between half-spring and half-mass contacts arises when $m_i=0$ and $1/k_i=0$. However, this OCI is an upper bound only for thermal conductance of systems in which $m_0$ vary arbitrarily and $1/k_i=0$ or reciprocally. When $k_0$ and $m_0$ can vary together, interferences enter to the picture and the upper bound is lost. 
Interface with varying mass $m_0=m_i+\frac{m_2}{2}$ and spring $\frac{1}{k_0}=\frac{1}{k_i}+\frac{1}{2k_1}$. The 1D chain modeled is obtained by combining the springs and masses when possible.}
	\label{fig2impint}
\end{figure}

Since the key to obtain an OCI is the use of symmetric blocks, one can imagine that an abrupt interface between contacts built with different types of block is also an OCI. In fact this is shown starting from a more general interface where two parameters $m_0$ and $k_0$ can be varied (Fig.~\ref{fig2impint}). Using the block concept to define the contacts and the impurities, phonon transmission is found to be
\begin{equation}
T=\frac{\Gamma_1^{hs}\Gamma_2^{hm}}{\left[\omega^2m_i+\frac{1}{k_i}\frac{\Gamma_1^{hs}}{2}\frac{\Gamma_2^{hm}}{2}\right]^2+\left[\left(\frac{\Gamma_1^{hs}}{2}+\frac{\Gamma_2^{hm}}{2}\right)-\omega^2\frac{m_i}{k_i}\frac{\Gamma_1^{hs}}{2}\right]^2}.
\label{equtsssm}
\end{equation}
From this system, two single junction interfaces can be defined by setting $k_i^{-1}$ (or $m_i$) to zero and letting  $m_0$ (or $k_0$) vary. The cancelations in the denominator of Eq.~\ref{equtsssm} expose an OCI when $m_i=0$ (or $k_i^{-1}=0$) which is an upper bound of thermal conductance for any other choice of $m_i$ (or $k_i^{-1}$). Similar to the cases in the last subsection, plotting $\Gamma_1^{hs}$ and $\Gamma_2^{hm}$ reflects some transmission characteristics (Fig.~\ref{figGammasm}c and \ref{figGammasm}d). If the system is studied as a whole and both impurities do not vanish, the transmission of the OCI is not necessarily an upper bound for every $\omega$. This does not contradict the previous definition of OCI because the interface allows two parameters to vary.

\subsection{Extrapolation Example: Beyond Single Interfaces}

\begin{figure}[tb]
\centering
\includegraphics[width=80mm]{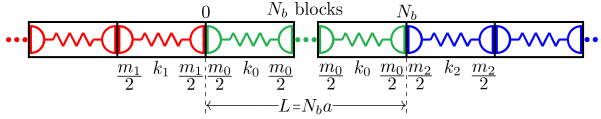}
\caption{System consisting of two mediums sandwiching a third one characterized by the broadenings  $\Gamma_1$, $\Gamma_2$ and $\Gamma_0$ respectively. Similar to the antireflection coating condition, transmission is unity when $L=\lambda/4$ and $\Gamma_0=\sqrt{\Gamma_1\Gamma_2}$.}
\label{fig3med}
\end{figure}

In subsection \ref{subseccvsd} we showed that OCIs generalize continuous interfaces to the discrete limit, which allows the extrapolation of known results between limits by changing interfaces with OCI and acoustic impedance $Z$ with broadening $\Gamma$. This analogy may provide a way to endow phonon engineering with existing design criteria from other engineering fields. For instance, broadband filter techniques from microwave engineering may be useful to engineer interfaces with maximum thermal conductance. As an example of the generalization, let us consider a system consisting of two mediums sandwiching a third one with impedances $Z_1$, $Z_2$ and $Z_0$ respectively. Recall that reflection is eliminated when the coupling medium has length of a quarter wavelength and impedance $Z_0=\sqrt{Z_1Z_2}$. Applying the extrapolation rules to the known solution \cite{Pain}, the system turns into Fig.~\ref{fig3med} and its transmission is given by (the result can also be obtained following the process in section \ref{sectransmission})

\begin{equation}
T=\frac{4\frac{\Gamma_1}{\Gamma_2}}{\left(\frac{\Gamma_1}{\Gamma_2}+1\right)^2\cos^2(q_0L)+\left(\frac{\Gamma_1}{\Gamma_0}+\frac{\Gamma_0}{\Gamma_2}\right)^2\sin^2(q_0L)},
\label{equtrans3m}
\end{equation}
with $\Gamma$'s defined according to the block choice in each particular region. Similar to the antireflection condition, transmission is unity when $L=\lambda/4$ and $\Gamma_0=\sqrt{\Gamma_1\Gamma_2}$. 

\begin{figure}[tb]
\centering
\includegraphics[width=50mm]{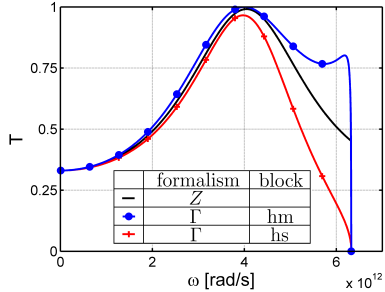}
\caption{Comparison of the transmission function predicted by Eq.~\ref{equtrans3m} (dotted line) and its counterpart with the long wavelength constraint, i.e. Eq~\ref{equtrans3m} replacing $\Gamma$ with $Z$ (solid line). The system consists only of half-mass blocks (Fig.~\ref{fig3med}) with $m_1=10^{-26}\ kg$, $m_2=10^{-24}\ kg$, $k_1=k_2=10\ N/m$ and $N_b=3$. $m_0$ and $k_0$ were chosen to guarantee $T=1$ @ $\omega=4\times10^{12}\ rad/s$, i.e. $m_0=1.1779\times 10^{-25}\ kg$ and $k_0=7.0338\ N/m$. The line with crosses represents the transmission of a system consisting only of half-spring blocks with the same parameters.}
\label{fig3medTvsw}
\end{figure}

Unlike the impedance formalism, i.e. Eq~\ref{equtrans3m} replacing $\Gamma$ with $Z$, the broadening formalism (Eq.~\ref{equtrans3m}) includes the effects of non linear dispersion and atomistic details. A comparison of the transmission predicted by both formalisms is shown in Fig.~\ref{fig3medTvsw}. Note that as the frequency increases and the non linearity of the dispersion becomes important, the transmission functions separate from each other. Also note that different atomistic details at the interface, defined by our block choice, generate different transmission functions.

\section{Conclusion}

This paper showed that the degree of mismatch at a single atom or bond interface depends on our ability to express the entire system solely in terms of building blocks on either side. Based on this concept, we argued that maximum thermal conductance happens when the mass or bond junctions are the {\it arithmetic} or {\it harmonic} mean of its neighbors respectively. Any deviation from those Optimally Coupled Interfaces (OCI) adds an extra barrier for heat carriers reducing the interfacial transmission and thermal conductance. We also showed that OCI and contact broadening ($\Gamma$) generalize continuous interfaces and acoustic impedance ($Z$) to the nonlinear phonon dispersion as well as the short-wavelength (atomic) limit. This generalization not only relates the continuum formalism with the discrete NEGF formalism, but also provides a way to extrapolate previous results based on acoustic impedance to OCI. This may allow us to use existing techniques from other engineering fields to phonon engineering. For instance, broadband filter techniques from microwave engineering may be useful to engineer interfaces with maximum thermal conductance.

\begin{acknowledgments}
C.A.P. thanks Professor Peter Arnold and Dr. John Duda for the useful discussions. C.A.P. and A.W.G. are grateful for the support from NSF-CAREER (QMHP 1028883). C.A.P., A.W.G. and P.E.H. are greatly appreciative for the funding from NSF-IDR (CBET 1134311). P.E.H. is thankful for the funding from the LDRD program office at Sandia National Laboratories. 
\end{acknowledgments}

\bibliography{bibpaper}

\end{document}